\documentclass[aps,pra,preprint,superscriptaddress, reprint,amsmath,amssymb,longbibliography]{revtex4-2}

\usepackage{graphicx} % Required for inserting images

\usepackage{xcolor}
\definecolor{darkblue}{rgb}{0,0,0.7}
\usepackage[colorlinks=true,linkcolor=darkblue,citecolor=darkblue,urlcolor=darkblue]{hyperref}
\usepackage{epsfig}
\usepackage{comment}
\usepackage{bm}
\usepackage{amsfonts}
\usepackage[utf8]{inputenc}
\usepackage{ulem}
\usepackage{textcomp} % µ in Roman: \textmu
\usepackage{siunitx}
\usepackage{afterpage}
\usepackage{soul}

\begin{document}
\title{\large\textbf{Inductive detection of inverse spin-orbit torques in magnetic heterostructures}}

\author{Misbah Yaqoob}
	\email{yaqoob@rptu.de}
	\affiliation{Fachbereich Physik and Landesforschungszentrum OPTIMAS, Rheinland-Pf\"{a}lzische Technische Universit\"{a}t Kaiserslautern-Landau, 67663 Kaiserslautern, Germany}

\author{Fabian Kammerbauer}
	\affiliation{Institut für Physik, Johannes Gutenberg-Universität Mainz, 55099 Mainz, Germany}

 \author{Tom G. Saunderson}
	\affiliation{Institut für Physik, Johannes Gutenberg-Universität Mainz, 55099 Mainz, Germany}

\author{Vitaliy~I.~Vasyuchka}
	\affiliation{Fachbereich Physik and Landesforschungszentrum OPTIMAS, Rheinland-Pf\"{a}lzische Technische Universit\"{a}t Kaiserslautern-Landau, 67663 Kaiserslautern, Germany}

 \author{Dongwook Go}
	\affiliation{Institut für Physik, Johannes Gutenberg-Universität Mainz, 55099 Mainz, Germany}

 \author{Hassan Al-Hamdo}
	\affiliation{Fachbereich Physik and Landesforschungszentrum OPTIMAS, Rheinland-Pf\"{a}lzische Technische Universit\"{a}t Kaiserslautern-Landau, 67663 Kaiserslautern, Germany}

 \author{Gerhard Jakob}
	\affiliation{Institut für Physik, Johannes Gutenberg-Universität Mainz, 55099 Mainz, Germany}

  \author{Yuriy Mokrousov}
	\affiliation{Institut für Physik, Johannes Gutenberg-Universität Mainz, 55099 Mainz, Germany}
    \affiliation{Peter Grünberg Institut and Institute for Advanced Simulation, Forschungszentrum Jülich and JARA, 52425 Jülich, Germany}
     
  \author{Mathias Kläui}
	\affiliation{Institut für Physik, Johannes Gutenberg-Universität Mainz, 55099 Mainz, Germany}

\author{Mathias Weiler}
    \email{weiler@physik.uni-kl.de}
    \affiliation{Fachbereich Physik and Landesforschungszentrum OPTIMAS, Rheinland-Pf\"{a}lzische Technische Universit\"{a}t Kaiserslautern-Landau, 67663 Kaiserslautern, Germany}

\date{\today}

\begin{abstract}
The manipulation of magnetization via magnetic torques is one of the most important phenomena in spintronics. In thin films, conventionally, a charge current flowing in a heavy metal is used to generate transverse spin currents and to exert torques on the magnetization of an adjacent ferromagnetic thin film layer. Here, in contrast to the typically employed heavy metals, we study spin-to-charge conversion in ferromagnetic heterostructures with large spin-orbit interaction that function as the torque-generating layers. In particular, we chose perpendicular magnetic anisotropy (PMA) multilayers [Co/Ni] and [Co/Pt] as the torque-generating layers and drive magnetization dynamics in metallic ferromagnetic thin film $\mathrm{Co_{20}Fe_{60}B_{20}}$ (CoFeB) layers with in-plane magnetic anisotropy (IMA). We investigate the spin dynamics driven by spin-orbit torque (SOT) and the concomitant charge current generation by the inverse SOT process using an inductive technique based on a vector network analyzer. In our experimental findings, we find that the SOTs generated by our multilayers are of a magnitude comparable to those produced by Pt, consistent with first-principles calculations. Furthermore, we noted a significant correlation between the SOT and the thickness of the CoFeB layer.
\end{abstract}

\maketitle
\section{INTRODUCTION}
\label{INTRODUCTION}
\begingroup
\setlength{\parskip}{1ex}
The spin-to-charge conversion facilitated by spin–orbit interaction, has emerged as a key focus in spintronics research due to its potential practical applications for energy-efficient and fast control of magnetization in spintronic devices~\cite{Hirsch:1999:SpinHallEffect,Kimura:2007:RoomTemperatureReversibleSpin,Saitoh:2006:SpinToCharge}. Conventionally, a heavy metal with strong spin-orbit coupling (SOC) is used for this purpose~\cite{Liu:2021:SpinOrbitTorques,Lee:2021:SpinOrbitTorques, Cha:2021:SpinorbitTorqueEfficiencya}. The interconversion of electrical charge currents and spin currents in ferromagnet/nonmagnet (FM/NM) multilayers connects electric currents to magnetic torques during the forward process, known as SOT~\cite{Rhodes:1961:7MicrosecondFerriteCorea, Divinskiy:2020:EffectsSpinOrbitTorque, Myers:1999:CurrentInducedSwitchingDomains, Parkin:1999:ExchangebiasedMagneticTunnel}. In the inverse process (iSOT)~\cite{Freimuth:2015:DirectInverseSpinorbit, Sun:2016:InverseSpinHall, Berger:2018:IndDetection}, they link magnetization dynamics with spin pumping and conversion of spin currents to the charge currents. The SOT and iSOT processes are thereby reciprocal~\cite{Freimuth:2015:DirectInverseSpinorbit}. The SOTs arising from the spin Hall effect (SHE) and Rashba–Edelstein effect in heavy nonmagnetic (NM) layers like Pt, $\beta$-Ta and W have been thoroughly examined and modeled~\cite{Czeschka:2011:ScalingBehaviorSpin, Du:2020:DisentanglementSpinOrbitTorques, Weiler:2014:DetectionDCInverse, Wang:2014:ScalingSpinHall}. Investigations into these torques have been conducted in both forward and inverse configurations. 

Another possible mechanism contributing to magnetic torques is the orbital Hall effect (OHE) that can also account for the extended dephasing length of angular momentum in ferromagnets~\cite{Liao:2022:EfficientOrbitalTorque,Hayashi:2023:ObservationLongrangeOrbital, Bose:2023:DetectionLongrangeOrbitalHall,Go:2018:IntrinsicSpinOrbital}. However, the efficiency of OHE generation varies among different magnetic materials~\cite{Tanaka:2008:IntrinsicSpinHall,Go:2020:TheoryCurrentinducedAngulara}, and experimentally distinguishing the OHE contribution from the SHE remains a significant challenge. A common experimental approach for spin-to-charge conversion involves injecting spin currents into a heavy metal from an adjacent ferromagnetic material and subsequently evaluating the resulting charge current or voltage. The spin current is often injected by exciting ferromagnetic resonance (FMR) and spin pumping~\cite{Tserkovnyak:2002:EnhancedGilbertDamping}. 

In this work, we employ a ferromagnetic material to generate SOTs using a method similar to that in Refs.~\cite{Baek:2018:SpinCurrentsSOTs,Humphries:2017:ObservationSpinorbitEffects}, which involves two ferromagnetic layers: one with perpendicular magnetization and the other with in-plane magnetization. This methodology has already been theoretically proposed in Ref.~\cite{Davidson:2020:Spincurrent}. We employ an ac inductive method~\cite{Berger:2018:IndDetection}, wherein an ac spin current is injected from a ferromagnetic layer into an adjacent layer by spin pumping. The ac spin current is converted into an ac electric current through the various spin-to-charge conversion processes. We explore spin-to-charge conversion in ferromagnetic heterostructures containing multilayers with PMA, specifically [Co/Ni] and [Co/Pt]. These structures induce SOTs in neighboring CoFeB thin films with IMA. We extract the spin dynamics and SOTs by applying vector network analyzer ferromagnetic resonance spectroscopy (VNA-FMR)~\cite{Tamaru:2018:VectorNetworkAnalyzer,Berger:2018:IndDetection, Meinert:2020:TechniquesSHE}. We have witnessed large SOTs from these ferromagnetic layers specially from [Co/Ni] multilayer which is comparable to what has already been observed in heavy metal Pt~\cite{Berger:2018:IndDetection,Du:2020:DisentanglementSpinOrbitTorques,Nguyen:2016:SpinTorqueStudy}.
\endgroup
\begin{figure*}[ht]
\centering
\includegraphics[width=6.29in]{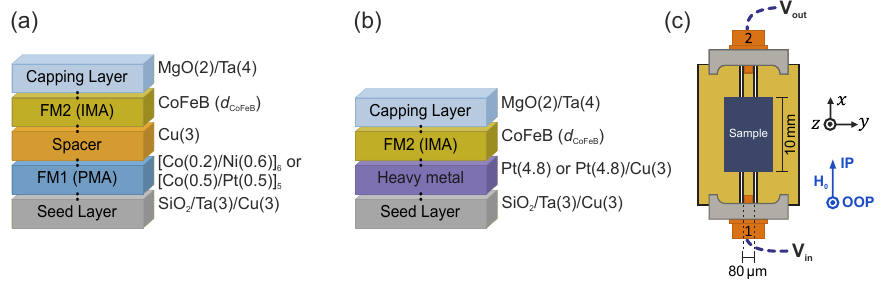}% Here is how to import EPS art
\caption{\label{fig:ML_stack_CPW}Sample growth order and measurement setup. (a) Sample stack with one ferromagnetic PMA layer and one ferromagnetic IMA layer separated by a nonmagnetic spacer. (b) Bilayer consisting of a heavy metal and an IMA ferromagnet. (c) Measurement setup with external magnetic field orientations for IP and OOP configurations and coplanar waveguide. The figure is not to scale for better visibility.}
\end{figure*}

\section{SAMPLE GROWTH}
\label{SAMPLEGROWTHANDSETUP}

To study the SOTs in all-ferromagnetic hybrids, we prepared several sample series with two different kind of sample stacks using sputter deposition. The first two sample series contain two layers of ferromagnets. One of the ferromagnet layers has PMA, and the other layer has IMA. These two layers of ferromagnets are separated by copper (Cu), a non-magnetic metal. For the PMA layers, we choose $[\text{Co(0.2)}/\text{Ni(0.6)}]_\text{6}$ and $[\text{Co(0.5)}/\text{Pt(0.5)}]_5$, where the numbers in parentheses indicate the thickness of the layer in nanometres. We prepared two series of metallic stacks for PMA samples consisting of substrate/Ta(3)/Cu(3)/[Co(0.2)/Ni(0.6)]$_6$/ Cu(3)/CoFeB($d_\text{CoFeB}$)/MgO(2)/Ta(4)\:\:and\:\:substrate/ Ta(3)/Cu(3)/[Co(0.5)/Pt(0.5)]$_5$/Cu(3)/CoFeB($d_\text{CoFeB}$) /MgO(2)/Ta(4) as shown in Fig.\,\ref{fig:ML_stack_CPW}(a). The third series of samples is a bilayer system containing a layer of heavy metal (Pt) and the same IMA ferromagnet (CoFeB) that we used for the PMA series. The bilayer series consists of substrate/Ta(3)/Cu(3)/Pt(4.8)/CoFeB($d_\text{CoFeB}$)/MgO(2)\\ /Ta(4) as shown in Fig.\,\ref{fig:ML_stack_CPW}(b). The thickness of Pt(4.8) is chosen to match the thickness of the PMA layer from the other series for a better comparison. This series was also investigated with the same spacer that we used for the first two PMA samples series, so the forth series contains substrate/Ta(3)/Cu(3)/Pt(4.8)/Cu(3)/CoFeB($d_\text{CoFeB}$)/\\MgO(2)/Ta(4) (sample stack not shown explicitly in Fig.\,\ref{fig:ML_stack_CPW}).   %         
\section{EXPERIMENTAL TECHNIQUE}
\label{EXPERIMENTALTECHNIQUE}
\begingroup
\setlength{\parskip}{1ex}
As demonstrated in Fig.\,\ref{fig:ML_stack_CPW}(c), the samples with \textit{l} = 10\,mm length are placed on a coplanar waveguide (CPW), facing down, with in-plane (IP) or out-of-plane (OOP) field $\mathbf{H}_0$ configuration. The actual size of the samples used is $6 \times \SI{10}{\milli \meter}^2$ with substrate thickness of \SI{525}{\micro \meter}. Further details about sample fabrication and substrate are provided in the Supplemental Material~\cite{SI}. This CPW has a characteristic impedance $Z_{0}$ = 50\,$\Omega$ and a signal line with width $W_\text{wg} = \SI{80}{\micro\meter}$. The CPW-sample assembly is placed between the pole shoes of an electromagnet which can produce a static magnetic field  up to $\mu_0 H_0\approx 2.5$\,T.  

For our experiments, we use the VNA-FMR technique to obtain magnetic-field and frequency-dependent $S_\text{21}$ parameters with complex amplitude. Quantitative evaluation of the $S_\text{21}$ data~\cite{Berger:2018:IndDetection,Berger:2018:SHEandSDL,Meinert:2020:TechniquesSHE,Shigematsu:2021:SpinChargeConversion} provides a measure of the spin-orbit torque (SOT)~\cite{Baek:2018:SpinCurrentsSOTs,Guimaraes:2020:SOTsGHzTHz,Shao:2021:RoadmapSOTs}. All measurements are carried out at room temperature. We measure the microwave transmission through the CPW loaded with the sample with the VNA as we sweep a dc magnetic field IP ($\mathbf{H}_\text{0}\parallel \hat{x}$)  or OOP ($\mathbf{H}_\text{0}\parallel \hat{z}$) [see Fig.\,\ref{fig:ML_stack_CPW}(c)] through the FMR condition of the CoFeB layer. To enhance the signal-to-noise ratio, we repeat the field sweeps and average the transmission data. Our technique takes advantage of the magnitude and phase of the signal to directly investigate the alternating charge currents generated by iSOT~\cite{Weiler:2014:PhaseSensitiveDetectionSpin, Wei:2014:SpinHallVoltagesa, Miao:2013:InverseSHE,Jiao:2013:SpinBackflowSHE}. We measure the change in microwave transmission

\begin{equation} 
\Delta S_{21} = - \frac{1}{2}{\left(\frac{i \omega L}{Z_\text{0}+i \omega L} \right)} \approx -\frac{ i \omega L}{2 Z_\text{0}}
\label{equ.Dealta_S21}
\end{equation}

where $\omega$ is the microwave frequency, that is related to the complex inductance $L$, where $L$ is the sample inductance from magnetization precession as well as ac current flow. Extraction of $L$ is critical for the analysis of our data, as it contains information on currents generated by odd symmetry SOTs that are commonly attributed to, e.g., spin pumping and the inverse spin Hall effect (iSHE)~\cite{Miao:2013:InverseSHE,Wahler:2016:ISHE} and also currents generated by even symmetry SOTs commonly attributed to, e.g., the inverse Rashba–Edelstein effect (iREE)~\cite{Zhang:2015:SpinPump,Yama:2023:TheoryIRE,Song:2021:IREeffect}. We note that both iSHE and iREE can generate SOTs of both symmetries and we thus refer to the symmetry of the torque (odd or even) and not to the microscopic mechanism when discussing our data.  
\endgroup
\section{DATA ANALYSIS}
\label{DATAANALYSIS}
\begingroup
\setlength{\parskip}{1ex}
\begin{figure}
\centering
\includegraphics{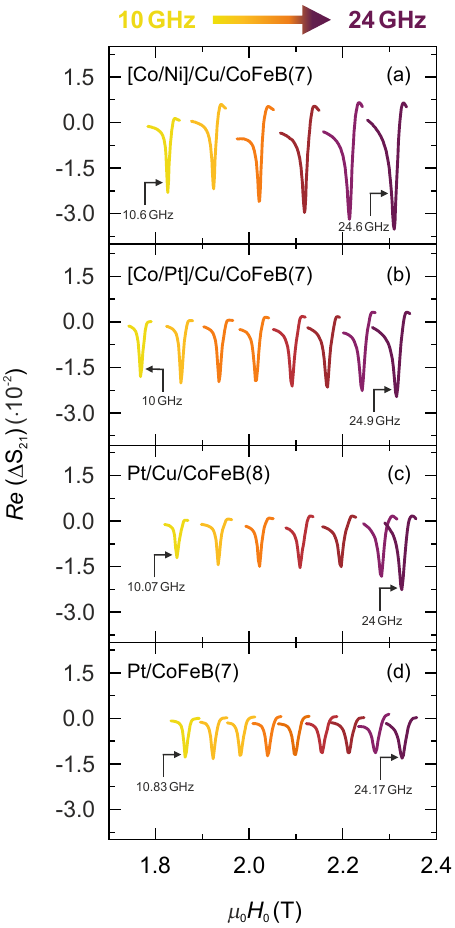}% Here is how to import EPS art
\caption{\label{fig:Re_deltaS21}Obtained $Re\,(\Delta S_{21}$) of FMR spectra for OOP configuration for four different sample types with excitation frequencies ranging from 10\,GHz to 24\,GHz. The first and last $Re\,(\Delta S_{21}$) spectra are labeled with their corresponding frequencies. (a) a PMA sample with [Co/Ni]/Cu/CoFeB. (b) a PMA sample with [Co/Pt]/Cu/CoFeB. (c) a sample with heavy metal and a spacer Pt/Cu/CoFeB. (d) a sample with a bilayer Pt/CoFeB. When $\textbf{H}_\text{0}$ meets the FMR condition, we observe $Re\,(\Delta S_{21}$) $\propto$ $Re\,(L)$. The alteration in the amplitude and line shape indicates the existence of frequency-dependent inductive components which is highest in [Co/Ni]/Cu/CoFeB.}  
\end{figure}
To study the effect of ferromagnet thickness on even symmetry and odd symmetry magnetic torques generated by a ferromagnetic layer or heavy metal layer with strong SOC~\cite{Sinova:2015:SHE,Saitoh:2006:SpinToCharge}, we investigate four different sample series as shown in Fig.\,\ref{fig:Re_deltaS21}. The sample growth sequence is indicated from left to right, with the substrate consistently positioned on the left. We obtain FMR spectra while varying the IP and OOP external magnetic field, $H_0$ with microwave frequencies ranging from 8\,GHz to 40\,GHz and 10\,GHz to 24\,GHz respectively. The VNA output power is set to 0 dBm, such that all measurements are carried out in the small-angle precession linear regime of the magnetization dynamics.
We begin by measuring $S_{21}$ VNA-FMR spectra and fit these data to

\begin{equation}
S_{21}(\omega , H_0) = S^0_{21}-iA e^{i \phi} \chi_\text{yy}(\omega , H_0)\;,
\label{equ.raw_S21}
\end{equation}

where $S^0_{21} =  C_0+C_1 H_0$ is the background originating from the frequency-dependent transmission of the setup with $C_0$ and $C_1$ as complex offset and slope corrections respectively to the background. Here, $\chi_\text{yy}(\omega , H_0)$ is the frequency-dependent magnetic susceptibility~\cite{Berger:2018:IndDetection, Weiler:2014:PhaseSensitiveDetectionSpin}. The magnitude of the signal is denoted by $A$ and $\phi$ is the raw phase of the signal. Therefore, a background-corrected signal $\Delta S_{21}$ is obtained as~\cite{Berger:2018:IndDetection}
\begin{equation}
\Delta S_{21}(\omega , H_0) = \frac{S_{21}(\omega , H_0)-S^0_{21}}{S^0_{21}}= -i\frac{Ae^{i\phi}\chi_\text{yy}(\omega , H_0)}{(C_0+C_1 H_0)}\;.
\label{equ.Corrected_Delta_S21}
\end{equation}
The background-corrected exemplary $Re\,(\Delta S_{21}$), OOP data for four different samples is shown in Fig.\,\ref{fig:Re_deltaS21}. For all samples, we observe a significant dependence of signal amplitude and phase on frequency. This indicates that the signal cannot be purely attributed to the dipolar inductance due to magnetization precession and concomitant voltage generation in the CPW by Faraday's law, which would always result in a pure dip-like signal shape in $Re\,(\Delta S_{21}$). In particular, a maximum change of lineshape is observed when we sweep through the entire frequency range for the [Co/Ni]/Cu/CoFeB sample. This indicates that this sample shows the highest spin-to-charge conversion efficiency in Fig.\,\ref{fig:Re_deltaS21}.
\begin{figure}[t]
\centering
\includegraphics{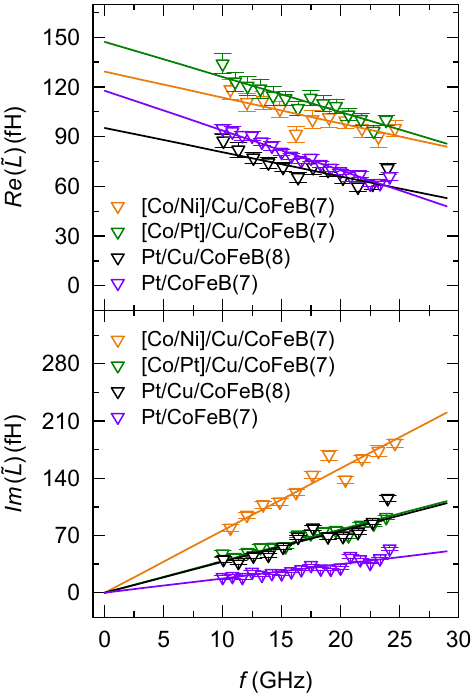}% Here is how to import EPS art
\caption{\label{fig:ReL_ImL}Real and imaginary inductances data (symbols) versus frequency together with the fits (solid lines) to  Eq.\,\eqref{equ.Re_Normalized_inductance} and Eq.\,\eqref{equ.Im_Normalized_inductance} for OOP configuration for same samples as in Fig.\,\ref{fig:Re_deltaS21}. (a) $Re\,(\tilde{L}$) for four different samples as a function of frequency. The slope of the fits gives the values of ($\sigma^\text{F}_\text{e} -\sigma^\text{SOT}_\text{e}$). (b) $Im\,(\tilde{L}$) for the same samples. Here the linear slope of the fits reflects $\sigma^\text{SOT}_\text{o}$.}
\end{figure}
For further data evaluation, we now concentrate on the complex-valued inductance $L$, which contains information about the magnitude and phase of the signal as shown in Fig.\,\ref{fig:Re_deltaS21}. To remove the impact of the CoFeB properties on $L$, we introduce the normalized inductance $\tilde{L}$. Using Eq.\,\eqref{equ.Dealta_S21} and Eq.\,\eqref{equ.Corrected_Delta_S21} and for $Z_0 \gg \omega L$ the normalized inductance $\tilde{L}=L/\chi_\text{yy}$ is calculated as
\begin{equation}
\tilde{L} (f)= \frac{ Ae^{i\phi}}{(C_0+C_1 H_0)}\frac{Z_0}{\pi f}\;.
\label{equ.Normalized_inductance}
\end{equation}   
Since $\tilde{L}$ is a complex quantity, we plot ${Re}\,(\tilde{L}$) and $Im\,(\tilde{L}$) as a function of frequency for selected samples in Fig.\,\ref{fig:ReL_ImL}. All samples have similar thicknesses of CoFeB. We define $\tilde{\sigma}_\text{FM1}$ or $\tilde{\sigma}_\text{NM}$$ $ = $(\sigma^\text{F}_{e}-\sigma^\text{SOT}_{e})+i\sigma^\text{SOT}_\text{o}$~\cite{Berger:2018:IndDetection} as the effective conductivity which contains both the magnitude and symmetry ($\sigma_{e}$ for even symmetry and $\sigma_{o}$ for odd symmetry with respect to time-reversal) of magnetic torques resulting from applied charge currents. Conversely, it also characterizes the alternating (ac) charge currents flowing within a sample in linear response to driven magnetization dynamics. The real part of $\tilde{\sigma}_\text{NM}$ consists of conditions both from Faraday effect ($\sigma^\text{F}_{e}$) and even symmetry SOTs ($\sigma^\text{SOT}_{e}$) while the imaginary part contains contributions from odd symmetry SOTs ($\sigma^\text{SOT}_\text{o}$). As shown in Fig.\,\ref{fig:ReL_ImL}, the measured complex-valued inductances scale linearly with frequency.  Following Berger \textit{et al.}~\cite{Berger:2018:IndDetection}, we fit our inductance data to  
\begin{equation}
\label{equ.Re_Normalized_inductance}
\begin{split}
Re (\tilde{L})  = \:&  \frac{\mu_0 l}{4} \biggr[ \frac{d_\text{FM2}}{W_\text{wg}} \eta^2 (z, W_\text{wg})
+\eta (z, W_\text{wg}) \\ 
& {\times \frac{L_\text{12}(z, W_\text{wg}, l) \hbar \omega} {\mu_0 l M_\text{s} e} }\left( \sigma^\text{F}_\text{e} -\sigma^\text{SOT}_\text{e} \right) \biggr]\;,
\end{split}
\end{equation}
\begin{equation} 
\label{equ.Im_Normalized_inductance}
\begin{split}
Im (\tilde{L})  = \:& \frac{\mu_0 l}{4} \biggr[ \eta (z, W_\text{wg}) 
\\ 
& {\times \frac{L_\text{12}(z, W_\text{wg}, l) \hbar \omega} {\mu_0 l M_\text{s} e}\sigma^\text{SOT}_\text{o}} \biggr] 
\end{split}
\end{equation}
to extract ($\sigma^\text{F}_\text{e} -\sigma^\text{SOT}_\text{e}$) and $\sigma^\text{SOT}_\text{o}$ from the slope of the fits.
Here, $\mu_\text{0}$ is the vacuum permeability, $M_\text{s}$ and $d_\text{FM2}$ are the saturation magnetization and thickness of the CoFeB layer, $\eta (z, W_\text{wg}) \equiv (2 / \pi) \tan^{-1}(W_\text{wg}/2z)$ is the spacing loss that reduces the signal induced into the CPW from currents in the sample due to the non-zero distance $z$ between the CPW and the sample~\cite{Berger:2018:IndDetection}. The value of $z$ for our experiments varies between about $\SI{3}{\micro \meter} \leq z \leq \SI{30}{\micro \meter}$. Exact values for $z$ for all measurements are shown in the Supplemental Material~\cite{SI}. The $L_\text{12}$ is the mutual inductance between the sample and CPW as described in Ref.~\cite{Berger:2018:IndDetection}.
\endgroup
%The $sgn( \hat{z} \cdot \hat{n})$ defines the sign of the term and depends on the directions of $\hat{z}$ and $\hat{n}$. The unit vector $\hat{z}$ for the coordinates along the z-axis is determined by the placement of samples on the CPW, with the reference point $z = 0$ established at the \textbf{FM}\stretchrel{$/$}{\textbf{NM}} interface, and $\hat{n}$ is defined as an interface normal that points to the FM. $L_{12}$ is the geometry-dependend mutual inductance between the sample and the CPW and is calculated as prescribed in Ref.~\cite{Berger:2018:IndDetection}.%

\section{RESULTS and DISCUSSION}
\label{RESULTSandDISCUSSION}

\subsection{\label{sec:exp-results} EXPERIMENTAL RESULTS}
\begingroup
\setlength{\parskip}{1ex}
\begin{figure*}[t]
\centering
\includegraphics[width=6.29in]{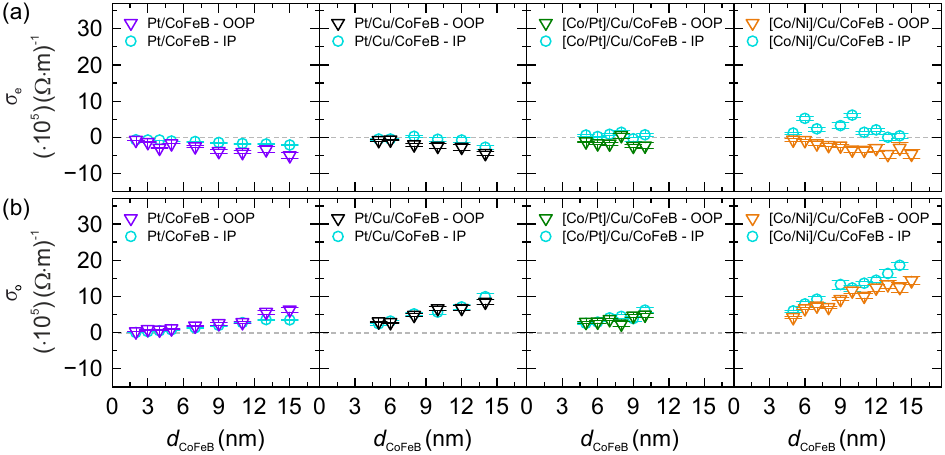}% Here is how to import EPS art
\caption{\label{fig:sigmaFL_sigmaDL} Even symmetry ($\sigma^\text{SOT}_\text{e}$) and odd symmetry ($\sigma^\text{SOT}_\text{o}$) SOTs conductivities as a function of thickness of CoFeB. (a) $\sigma^\text{SOT}_\text{e}$ for four different samples in the IP (circles) and OOP (triangles) configuration with CoFeB thickness ranging from 2\,nm $-$ 15\,nm. (b) $\sigma^\text{SOT}_\text{o}$ of the same samples with similar thicknesses of CoFeB in the IP (circles) and OOP (triangles) configuration. $\sigma^\text{SOT}_\text{o}$ is lowest for Pt/CoFeB sample series and highest for [Co/Ni]/Cu/CoFeB series.}
\end{figure*} 
The effective conductivities $\sigma^\text{SOT}_\text{e}$ and $\sigma^\text{SOT}_\text{o}$ for four different series of samples in IP and OOP are shown in Fig.\,\ref{fig:sigmaFL_sigmaDL}. Panel (a) summarizes the results of $\sigma^\text{SOT}_\text{e}$. It is important to note that here we did not separate the Faraday conductivity $\sigma^\text{F}_\text{e}$ contribution from $\sigma^\text{SOT}_\text{e}$ (as from Eq.\,\eqref{equ.Re_Normalized_inductance}) because we assume that it is identical for all samples due to the fact that the thicknesses/material of seed layer, spacer and capping layer are same for all samples. For all samples, the values of $\sigma^\text{F}_\text{e}$ are comparatively small, which we attribute to small torques with even symmetry (iREE or Oersted fields) in all investigated samples. Panel (b) represents the extracted $\sigma^\text{SOT}_\text{o}$  of the same samples. We find $\sigma^\text{SOT}_\text{o}> \sigma^\text{SOT}_\text{e}$ for all samples. The sample series with the [Co/Ni] multilayer shows the overall highest $\sigma^\text{SOT}_\text{o}$. For the sample with $d_\mathrm{CoFeB}=\SI{5}{\nano\meter}$ in this sample series we obtain $\sigma^\text{SOT}_\text{o} = 4.19 \pm 0.37 \times 10^5\,(\Omega\cdot \mathrm{m})^{-1}$.

%In our experiment, we can establish a connection between the effective conductivities ($\sigma^\text{SOT}_\text{e}$ and $\sigma^\text{SOT}_\text{o}$ ) and the microscopic spin-charge conversion~\cite{Shigematsu:2021:SpinChargeConversion} in the region where we have a very thin CoFeB layer (up to $\sim$ 6\,nm). We believe that, up to 6\,nm, where we observe the enhancement of Gilbert damping as shown in SI, the torques are generated due to iSHE. Here, the connection is made under the assumption that the damping-like iSOT arises from iSHE, while the field-like iSOT originates from iREE. We consider that the Faraday contribution $\sigma^\text{F}_\text{e}$ to the alternating charge currents in the normal metal or FM1 is related to the characteristics of the sample.
%    
Here, we can compare our measured values of $\sigma^\text{SOT}_\text{o}$ with those obtained by other groups, employing either different or similar techniques. Berger \textit{et al.}~\cite{Berger:2018:IndDetection} used the same inductive technique and reported a value of $\ 2.4 \times 10^5\,(\Omega\cdot \mathrm{m})^{-1}$ for 
 Ta(1.5)/Py(3.5)/Pt(6)/Ta(3) and $\approx 2.1 \times 10^5\,(\Omega\cdot \mathrm{m})^{-1}$  for $\mathrm{Ta(1.5)/Py(3.5)/Pt(5)/Ta(3)}$~\cite{Berger:2018:SHEandSDL}. Hibino \textit{et al.} used Co-Ni-B alloy films and measured via harmonic Hall measurements for $ \mathrm{Ta(2)/(Co_\text{x}Ni_\text{1-x})_\text{80}B_\text{20}(10)/ Ta(5)}$~\cite{Hibino:2020:LargeSpinOrbitTorqueEfficiency}, they find $\approx 1.0 \times 10^5\,(\Omega\cdot \mathrm{m})^{-1}$. Using the harmonic Hall method, Garello \textit{et al.}~\cite{Garello:2013:SymmetryMagnitudeSpin} for $\mathrm{AlO_\textbf{x}(2)/Co(0.6)/Pt(3)}$ and Nguyen \textit{et al.}~\cite{Nguyen:2016:SpinTorqueStudy} for $\mathrm{Ta(1)/Pt(\textit{t}_\text{Pt})/Co(1)/MgO(2)/Ta(1)}$ have reported values $\approx 5.5 \times 10^5\,(\Omega\cdot \mathrm{m})^{-1}$ and $\approx 2.5 \times 10^5\,(\Omega\cdot \mathrm{m})^{-1}$ respectively. All of these reported values are comparable to our findings. We thus believe that, up to a CoFeB thickness of about 6\,nm, where we observe the enhancement of Gilbert damping as shown in the Supplemental Material~\cite{SI}, the torques in all samples are compatible with an iSHE mechanism. 

In all sample series, the values of $\sigma^\text{SOT}_\text{o}$ continue to increase with the thickness of CoFeB even after 6\,nm. This is not expected from a conventional iSHE mechanism of charge generation which should show an eventual saturation of $\sigma^\text{SOT}_\text{o}$.
%After $6\,$nm we do not observe an enhancement in Gilbert damping in the system, but remarkably $\sigma^\text{SOT}_\text{o}$ continue to increase as we go further with the higher thicknesses of CoFeB.
A possible reason for this behavior could be a self-induced SOT~\cite{Aoki:2022:AnomalousSignInversion,Seki:2021:SpinorbitTorqueNiFe} in the CoFeB layer. Du \textit{et al.}~\cite{Du:2021:OriginSpinOrbita} have reported in their work that due to the self-induced SOT, the effective spin Hall conductivity $\sigma_\mathrm{xx}^s (\sigma^\text{SOT}_\text{o})$ increase linearly with the thickness of the ferromagnetic $\mathrm{Co_{20}Fe_{60}B_{20}}$ layer well beyond \SI{10}{\nano\meter} thickness,  which is quite similar to what we have observed in our experiment. Note that this is different from the trend in iSHE voltage with thickness of the ferromagnet such as reported in the work of  Nakayama \textit{et al.}~\cite{Nakayama:2012:GeometryDependenceInversea}, as the iSHE voltage is impacted by the line broadening due to spin pumping. In our evaluation, we use the normalized inductance, such that our $\sigma^\text{SOT}_\text{o}$ are not impacted by the dynamic properties of the ferromagnet -in particular the linewidth of the ferromagnetic resonance. 
%In  where they have already investigated the evolution of iSHE on the thickness of the ferromagnetic layer in a $\mathrm{Ni_{81}Fe_{19}/Pt}$ bilayer film.
%Considering a parallel resistor model, $V_\text{ISHE}= R R_\text{F} I_\text{c}/(R \, +\, R_\text{F})$, where $V_\text{ISHE}$ is the inverse spin Hall voltage and $R$ and $R_\text{F}$ are the resistance of normal metal layer and resistance of ferromagnetic layer respectively, they observed that the charge current, $I_\text{c}$ increases linearly with the thickness of ferromagnetic layer up to a saturation value. The $I_\text{c} \propto \sigma^\text{SOT}$ as $J = \hbar/2e ( \sigma^\text{SOT} \omega )$~\cite{Berger:2018:IndDetection}, where $J$ is the charge current density. It implies that, our $\sigma^\text{SOT}_\text{o}$ values are still below saturation value but by further increasing the thickness of CoFeB layer, we will achieve the saturation state for $\sigma^\text{SOT}_\text{o}$. In this region (below saturation point) we also do not need to do shunt correction~\cite{Berger:2018:SHEandSDL} because the sheet resistance, $R_\square$ of all of our samples is higher than 100\,$\Omega/ \square$ which is higher than the CPW characteristic impedance of 50\,$\Omega$. At this point, we must say that we need further investigations to find the real origin of this torque and we emphasize that a reliable theoretical model might be established to support the experimental results.

Comparing the results of  bilayer Pt/CoFeB with a trilayer Pt/Cu/CoFeB, we observe higher values of $\sigma^\text{SOT}_\text{o}$ in the trilayer sample series. We believe that a nonlocal direct spin current generation mechanism~\cite{Amin:2023:DirectIndirectSpin} is responsible for the increased torques observed in trilayers. This mechanism permits spin components that are symmetry-forbidden in bilayers to contribute to torques with odd symmetry. Another intriguing factor that may significantly contribute is the interfaces between Cu/[Co/Ni] and Cu/[Co/Pt], respectively. These interfaces may exhibit unique electronic states capable of generating substantial spin currents, in accordance with the concept of \textit{interface-generated spin currents}~\cite{Amin:2018:InterfaceGeneratedSpinCurrentsa}.
%So far, Pt has been studied extensively in the context of SOTs and found to be a promising material with large SOTs. On the other hand, there are groups who have utilized ferromagnetic materials with strong spin-orbit coupling (SOC) instead of heavy metals to generate such torques ~\cite{Humphries:2017:ObservationSpinorbitEffects,Baek:2018:SpinCurrentsSOTs}. We can contrast our recorded values of $\sigma^\text{SOT}_\text{o}$ with those obtained by other groups employing varying methodologies and materials.
%The reported value for pure Pt by M. Meinert \textit{et al},. ~\cite{Meinert:2020:TechniquesSHE} have reported a value of
\endgroup
\subsection{\label{sec:level2} FIRST-PRINCIPLES CALCULATIONS}
\begingroup
\setlength{\parskip}{1ex}
\begin{figure}[b]
\centering
\includegraphics{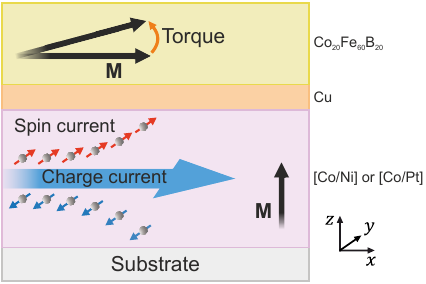}% Here is how to import EPS art
\caption{\label{fig:proces_schematics} Schematics of the process for DFT calculations. The [Co/Ni] or [Co/Pt] multilayer has PMA and CoFeB layer has IMA. The charge current in multilayers generates the spin current which transfers a torque to the CoFeB magnetization ($\mathbf{M}$).}
\end{figure}
While our experiments cannot reveal the microscopic origin of the significant SOTs that we observe for thick CoFeB films, the large SOTs that are generated in particular in the [Co/Ni]-based samples already in the limit of thin CoFeB films can be compared to expectations from first-principles calculations. We consider bulk multilayers of [Co/Ni] and [Co/Pt] and take the compositions of each material according to our experimental specifications. The schematics of this process is depicted in Fig.\,\ref{fig:proces_schematics} for clarity.  

We compare our density functional theory calculation results with the typical value of the spin Hall conductivity of Pt~\cite{Guo:2008:IntrinsicSpinHall}. The results are presented in Fig.\,\ref{fig:DFT_plot_SHE_all} and details about the calculations are provided in the Supplemental Material~\cite{SI}. The Fermi energy is varied by adding or removing electrons in the calculation, assuming that the band structure remains the same as the one obtained at the true Fermi energy, $E_\text{F}^\text{true}$ where the number of electrons equals the number of total nuclear charges. In our case, at the interface with another material such as Cu, charge transfer may happen, which would change the Fermi energy slightly therefore, we take ($E_\text{F}-E_\text{F}^\text{true}$) instead of $E_\text{F}^\text{true}$ only. Usually, the variation of various physical quantities as a function of the Fermi energy is checked, in case the quantity sensitively depends on the doping level.
\begin{figure}[t!]
\centering
\includegraphics{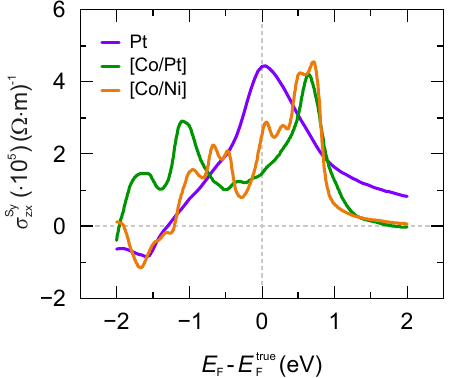}% Here is how to import EPS art
\caption{\label{fig:DFT_plot_SHE_all} Calculated values of the SHE conductivities for [Co/Ni] and [Co/Pt], shown in  orange and green lines, respectively. The result for Pt is also shown as a reference (purple line). Note that the SHE conductivities of [Co/Ni] and [Co/Pt] are sizable and comparable to that of Pt.}
\end{figure}
For Pt, at the Fermi energy, the value of SHE conductivity $\approx 4.43 \times 10^{5}\,(\Omega\cdot \mathrm{m})^{-1}$. For [Co/Ni] and [Co/Pt], the maximum values of the SHE conductivities are similar to that of Pt, which appear  $\sim$\,0.7\,eV above the Fermi energy. Even at the true Fermi energy, the SHE conductivities of [Co/Ni] and [Co/Pt] are $4.55 \times 10^{5}\,(\Omega\cdot \mathrm{m})^{-1}$ and $4.19 \times 10^{5}\,(\Omega\cdot \mathrm{m})^{-1}$ respectively, which are sizable. These first-principles calculations demonstrate that [Co/Ni] and [Co/Pt] multilayers can generate SHE-driven torques that are comparable to those in Pt, in good agreement with our experimental data.
\\
\section{CONCLUSIONS}
\label{CONCLUSIONS}
In summary, we conducted inductive ac measurements of a bilayer Pt/CoFeB and trilayer Pt/Cu/CoFeB, [Co/Ni]/Cu/CoFeB and [Co/Pt]/Cu/CoFeB samples with different thicknesses of CoFeB using phase sensitive VNA-FMR. We did not observe significant even symmetry spin-orbit torque conductivity, $\sigma^\text{SOT}_\text{e}$ in these samples. However, we observe higher values of odd symmetry spin-orbit torque conductivity, $\sigma^\text{SOT}_\text{o}$ where [Co/Ni] multilayer shows the highest value. The odd symmetry spin-orbit torque conductivity increases linearly with the thickness of CoFeB, indicating that a pure SHE and REE mechanisms are not sufficient to explain our dataset. In the limit of small CoFeB thicknesses our findings agree with previous results and align with our first principles calculations. The large torques that we revealed in [Co/Ni]-based samples may be relevant for characterization and optimization of future spintronic devices.
\begin{acknowledgments}
%We acknowledge the fruitful discussion with co-authors from Johannes Gutenberg-Universität Mainz and Peter Grünberg Institut and Institute for Advanced Simulation, Forschungszentrum Jülich and JARA. 
This work was financially supported by Deutsche Forschungsgemeinschaft (DFG, German Research Foundation) via TRR 173 “Spin+X” 268565370 (projects B13, A01, B02 and A11).
\end{acknowledgments}
\endgroup

%\nocite{*}

%\bibliography{revised}% Produces the bibliography via BibTeX.


\end{document}